\documentclass[ amsmath, amssymb, aps, 12pt,superscriptaddress]{revtex4-1}

\usepackage{graphicx}
\usepackage{dcolumn}
\usepackage{bm}
\usepackage{hyperref}
\usepackage{xcolor, soul}

\begin{document}

\title{Diamond-based optical vector magnetometer}

\author{Charlie Oncebay Segura}
\affiliation{ Instituto de Física de São Carlos, Universidade de São Paulo (IFSC-USP),
\\ Caixa Postal 369, CEP 13560-970, São Carlos, SP, Brazil. \\ ( srmuniz@ifsc.usp.br -- https://orcid.org/0000-0002-8753-4659 ) }
\affiliation{Facultad de Ciencias, Universidad Nacional de Ingeniería, Lima, Peru.}

\author{Sérgio Ricardo Muniz} 
\affiliation{ Instituto de Física de São Carlos, Universidade de São Paulo (IFSC-USP),
\\ Caixa Postal 369, CEP 13560-970, São Carlos, SP, Brazil. \\ ( srmuniz@ifsc.usp.br -- https://orcid.org/0000-0002-8753-4659 ) }

\date{31-May-2021}

\begin{abstract}
\vspace{7mm}
We describe here* the construction and characterization of a high-resolution optical magnetometer to measure the full vector magnetic field on an ultrathin layer near the surface of the device. This solid-state device is based on quantum sensors created by a layer of nitrogen-vacancy (NV) centers less than 20 nm below the surface of an ultrapure diamond. This ensemble of nanosensors provides a versatile device capable of mapping magnetic fields and surface current densities with a sub-micrometer resolution and high sensitivity, making it suitable for many applications. Here, we show a custom-built prototype to demonstrate an operating proof-of-concept device. \\
(*) \emph{Paper presented at the Conference \href{https://doi.org/10.1109/SBFotonIOPC50774.2021.9461950}{SBFoton-IOPC-2021}.}
\end{abstract}
\maketitle

\section{Introduction}
The nitrogen-vacancy (NV) center in diamond is a promising platform for many applications in quantum technologies. Among these, quantum sensing  
and, mainly, magnetometry are the most promising \cite{Taylor,Steinert1,Rondin1,Tetienne_graphene}. 

Diamond itself has remarkable material properties, making it well-suited for building microscopic field-deployable devices in a wide variety of environments: from zero to high temperatures, in a wide range of pressures, and even harsh chemical conditions. Since diamond is biocompatible, it can also be used for sensing biological samples \cite{Davis1}, even inside cells (using nanodiamonds), and sensitive organic materials not compatible with other methods capable of nanometric resolution magnetometry.

The use of NV center (NVC) for magnetometry applications is usually discussed in two main contexts: 1) either as a single NVC scanning probe \cite{Horowitz1,Appel1}, or 2) as an ensemble of NVCs \cite{Rondin1,Davis1} for magnetic field sensing and imaging. Here we describe a proof-of-principle application aiming to study the electronic properties of 2D materials \cite{Tetienne_graphene, Charlie_PhD-Thesis}, using an ensemble of NV centers engineered in an ultrathin layer near the surface of a bulk ultrapure diamond. This device combines fluorescence microscopy and optically detected magnetic resonance (ODMR) with electronic spin magnetic resonance spectroscopy to build a vector magnetometer capable of directly imaging all the vector components on the NVC plane. To demonstrate it, we used a quasi-2D model circuit to produce a current near the surface of the device and measured the full B-field created by the electric current. Using the vector field information, we reconstructed the current density vector that generated the magnetic field.

\section{Magnetometry with NV centers} \label{theory}

\subsection{NVC-based optical magnetometry}
The atom-like energy level structure of the NV center in the diamond lattice \cite{Acosta2} makes it a vector magnetometer of sub-nanometer resolution \cite{Rondin1,Tetienne_graphene,Steinert1}. The NV center is a point defect in the diamond crystal, consisting of a substitutional nitrogen atom combined with an adjacent carbon vacancy. It is a color center that absorbs photons in the visible range of 500 – 620 nm and emits photons in a broad range of 632–800 nm. 
The photoluminescence spectrum has two zero phonon lines (ZPL) \cite{DOHERTY1,Charlie_PhD-Thesis}. One line, at 575 nm, is due to neutral centers $\mbox{NV}^0$. The second ZPL, at 637 nm, belongs to $\mbox{NV}^-$, the negatively charged centers. Here, we are interested only in the $\mbox{NV}^-$, simplifying the notation to NV, meaning the negatively charged centers.

\begin{figure}[b]

\centerline{\includegraphics[scale=0.7]{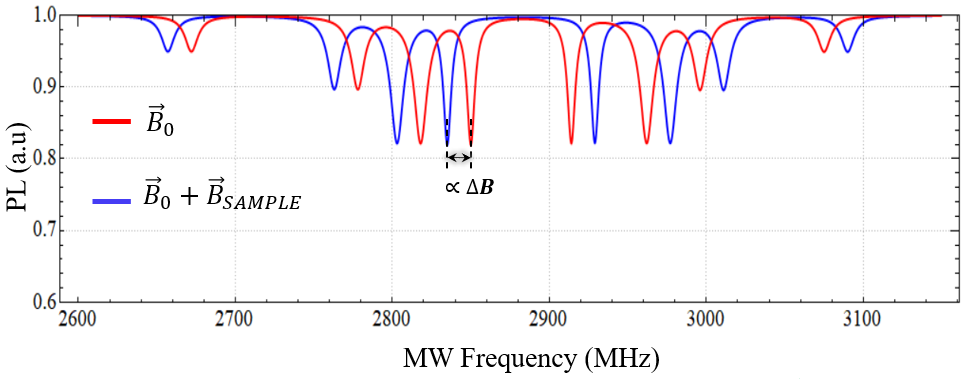}}
\caption{Simulated ODMR spectra for different magnetic fields, showing Zeeman shifts due the different directions of the NV axis.\\ \\}
\label{Zeeman}
\end{figure}

References \cite{Yuchen, Rogers1} propose a model where NV centers have a triplet ground state ($m_s=0, \pm1$) with a zero-field splitting $D_{GS} = 2.87$ GHz, a triplet excited state with the zero-field splitting of 1.42 GHz, and shelving states involved in inter-system crossing. Applying a constant magnetic field $\vec{B}$ is possible to remove the degeneracy of the $m_s= \pm1$ states and the evolution of the relevant states is governed by the (simplified) effective Hamiltonian
\begin{equation}
H= D_{GS}S_z^2+E(S_x^2+S_y^2)+\gamma_e\,\vec{B}.\vec{S},
\label{Hamiltonian}
\end{equation}
where the parameter $D_{GS}$ is the axial zero field splitting, which is also sensitive to temperature \cite{AcostaTemp}, $E$ is the transversal zero field splitting, $\gamma_e$ is the electron gyromagnetic ratio, and $\vec{B}$ is the external magnetic field. The spin vector $\vec{S}=(S_x,S_y,S_z)$ represents the electron spin, composed by the spin operators, here, given by $S_i = \sigma_i$, where $\sigma_i$ are the Pauli matrices. Thus, the applied magnetic field can be determined by the Zeeman shifts of the electron spin levels, as illustrated in Fig. \ref{Zeeman}.

\begin{figure}
\centerline{\includegraphics[scale=0.75]{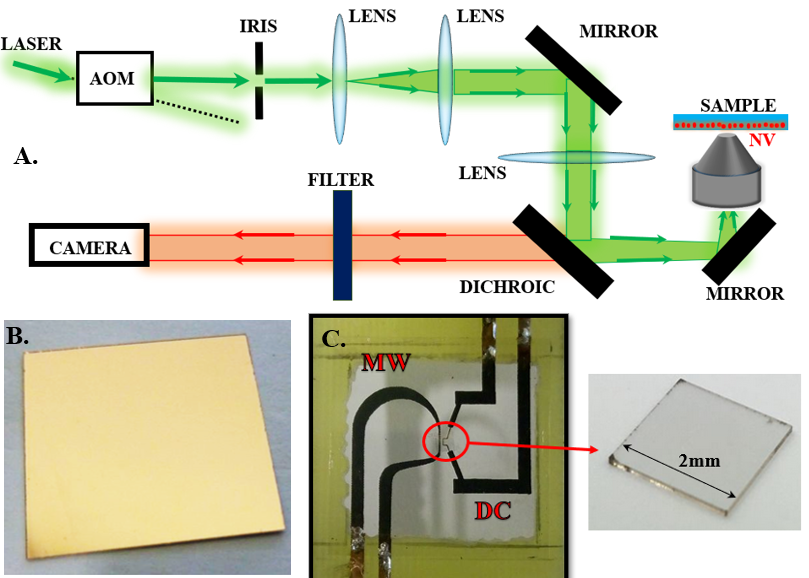}}
\caption{(A) Schematic of the apparatus, with wide-field fluorescence microscope, imaging system and excitation laser at $532$ nm. (B) Thin glass plate with deposited layer of copper. (C) Diamond plate is glued on the  glass plate with etched circuits for a single-loop microwave antenna (MW) and a straight section of copper wire for the DC circuit \cite{Charlie_PhD-Thesis} .}

\label{setup}
\end{figure}

The optimum sensitivity of a magnetometer base on a  ensemble of NV centers to measure DC magnetic fields depends on the intensity profile of NV resonances in the ODMR spectrum following the relation \cite{Dreau}
\begin{equation}
\eta= 0.77 \frac{h}{g\mu_B}\frac{\Delta f}{\alpha\sqrt{R}}
\label{sensitivity}
\end{equation}
where $\Delta f$ is the linewidth and $\alpha$ is the line contrast. Note that one may increase $\alpha$ by increasing the power of the MW excitation at the expense of increasing $\Delta f$ due to power broadening. The term $\sqrt{R}$ corresponds to the shot-noise, at a photon rate $R$. Using this relation, one can determine the best set of parameters to achieve the desired sensitivity of the magnetic field.

\section{Experimental setup}
The sensor used in this study was engineered from a type IIa ultrapure diamond plate. The plate was thinned and repolished until obtaining a chip of size  $2\times2\times 0.1\,\mbox{mm}^3$. A thin layer of  NV centers was created near the top surface by irradiating it with $^{15}\mbox{N}^+$ ions at $5$ keV and $1\times10^{13}\, \mbox{ions/cm}^{2}$ beam density. After implantation, the sample was annealed at $800\, ^o$C, in vacuum, for a couple of hours, to allow the migration and trapping of the vacancies to the implanted nitrogen atoms. These conditions create a layer of NVCs at around $8$ to $16$ nm below the surface at an estimated density of $10^3\, \mbox{NV}/\mu\mbox{m}^2$ \cite{Tetienne_graphene,Haque-Sumaiya,Pezzagna,Jamieson}. 

To excite the sample and collect its fluorescence, we built a custom wide-field fluorescence microscope, sketched in Fig. \ref{setup}. After the acousto-optical modulator (AOM) and the iris, the laser is expanded and collimated with two lenses to obtain a beam diameter of $\sim 10\,\text{mm}$. 
The collimated beam goes to a dichroic mirror (SemRock, Di02-R561-25x36) and is focused using a lens ($f=30 \text{ cm}$) near the back aperture of the microscope objective (Zeiss, $50\times$, 0.95 NA) to adjust the size of the illuminated area on the sample. The position control of the sample was provided by a nanopositioner (Thorlabs, NanoMax 300), enabling independent translations along three axes. The NV fluorescence is collected by the same microscope objective used to excite the sample and filtered with the dichroic mirror and long-pass filter at 550 nm (FELH0550). The photoluminescence is detected with a simple CCD camera (PointGrey, FL3-FW-03S1M-C).

\begin{figure}[htbp]
\centerline{\includegraphics[scale=1.3]{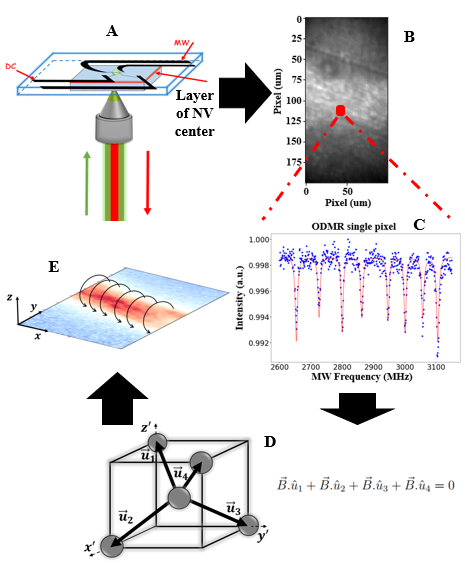}}
\caption{
Protocol used to obtain the magnetic field with the NV center considering the diamond chip has any crystallographic orientation \cite{Charlie_PhD-Thesis}.
\label{protocolo}
}
\end{figure}

The microwave (MW) and the DC wires were prepared in-house using a copper film (100 nm thickness) deposited by sputtering on the surface of a thin glass plate. We designed the MW and DC lines and printed them on photo (glossy) paper, applying the printed pattern onto the adequately cleaned copper surface. Later, we put the print-transferred glass plate into a chemical etching solution (ferric chloride solution) for about 30 seconds to remove the excess copper film. Finally, we cleaned and soldered the thin etched lines to thicker copper lines, etched on a usual circuit board.

The microwave excitation was provided by a programable signal generator (Standford Research, SG384), connected to a fast switch (CMCS0947A-C2) to control the MW pulse and an amplifier (Mini-circuits, ZHL-16W-43-S). The timing sequences were provided by a digital card (SpinCore-PulseBlaster, PBESR-PRO-300), producing pulses with 3 ns resolution to control the laser and microwave excitation, and the camera trigger. Custom Python codes  \cite{Charlie_PhD-Thesis} were used to control the system, and to save and analyze the images.

\section{Methods and Results}
The protocol used to determine the vector B-field is shown in Fig. \ref{protocolo}, where the diamond chip was glued onto a glass coverslip with the MW and the test DC-circuits. The ODMR spectrum comprises 64 images taken at each frequency, scanned using 0.5 MHz steps, to build the entire spectrum for each pixel in the fluorescing area. The spectra at each point (pixel) is used to determine the vector magnetic field at that location. 

\begin{figure}[tb]
\centerline{\includegraphics[scale=0.7]{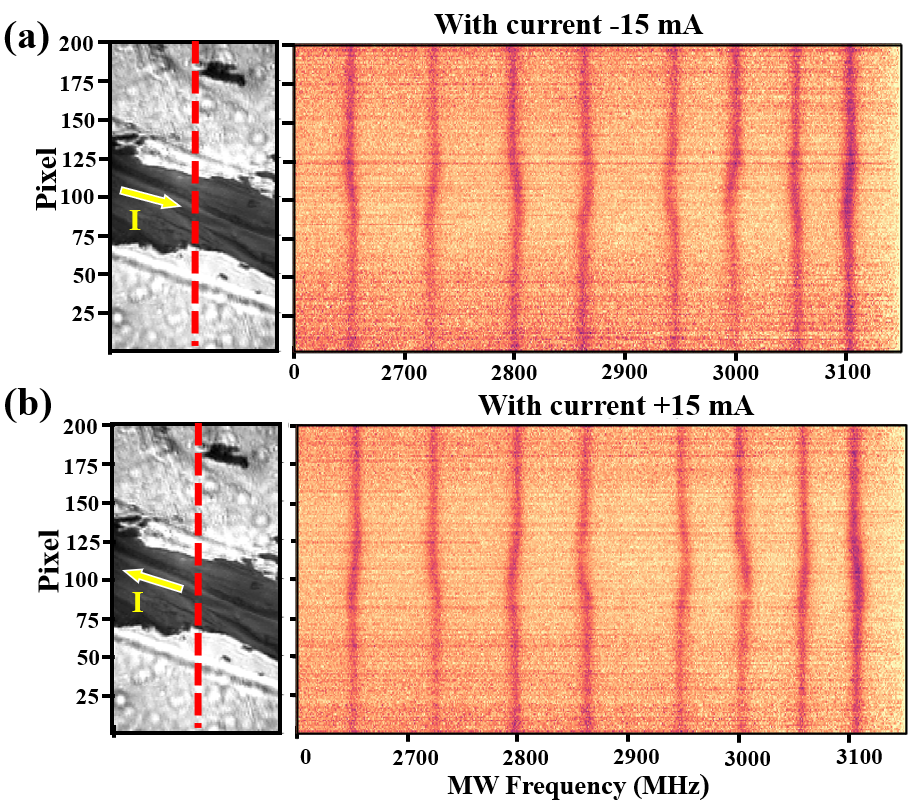}}
\caption{(a)-(b) ODMR spectra measured at points on the highlighted red dashed-line, for currents $\pm$15 mA \cite{Charlie_PhD-Thesis}. The microscopy image, showing a section of the wire on the left, is $100\times200$ pixels (1 pixel = 0.7 $\mu$m).
}
\vspace{-1mm}
\label{wire}
\end{figure}

We use a permanent magnet to apply a static bias field ($\sim10$ mT), splitting the degenerate spin states into eight lines corresponding to four pairs of spin resonances (2 for each direction of the NV-axis).  The visibility of the lines depends on the orientation of the magnet. In addition to the bias field, when an electric current passes through the wire, we observe frequency shifts in the ODMR spectrum due to the magnetic field produced around the DC wire, see Fig. \ref{wire}(c) and \ref{wire}(d). Note that the frequency shifts in the ODMR spectrum, caused by the current in the wire, are identical and symmetrically displaced for both directions of the current. To increase the signal-to-noise ratio, we record images with for both signs of the current, performing a differential measurement. We recorded three sequences of images: one without current and two with opposite signs of currents, as shown in Fig. \ref{wire}. This procedure allows to minimize common mode noise and offsets in the current, and to account for background and bias fields.

\begin{figure}[b]
\centerline{\includegraphics[scale=0.8]{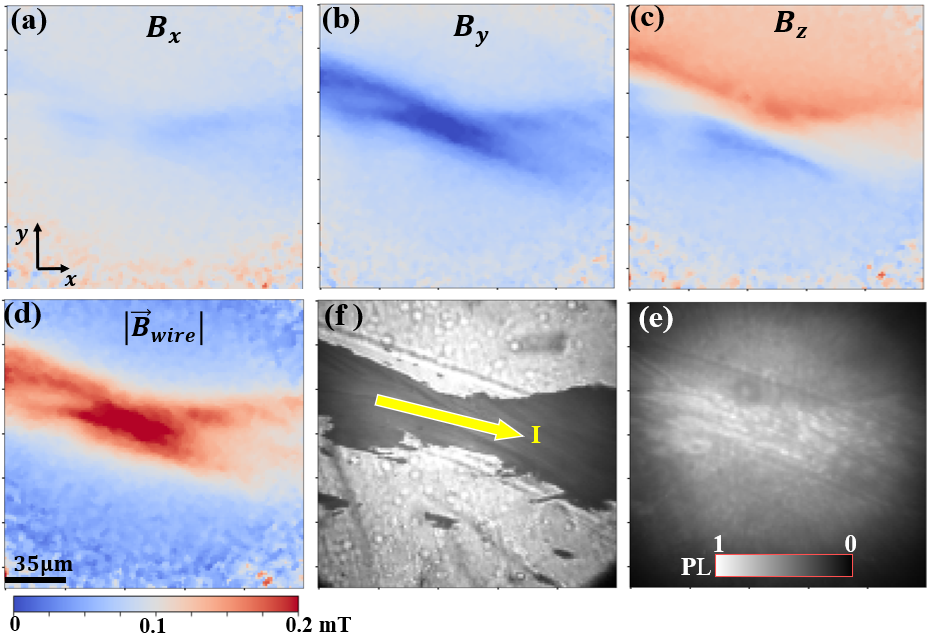}}
\caption{(a)--(c) Components of the magnetic field produced by a current $I=15\,\operatorname{mA}$. (d) Magnitude of the field  at the NV layer. (e) Microscopy image for a small section of the DC wire. (f) Direct normalized image of the photoluminescence (PL), during continuous laser excitation \cite{Charlie_PhD-Thesis}.}
\label{Bwire}
\end{figure}

To reconstruct the vector magnetic field at each pixel, we used a multi-Lorentzian fit on the ODMR spectra to extract the eight resonance frequencies ($\nu^{i}_{\pm}$). 
The frequencies satisfy $\nu^{i}_{\pm}=D_{GS}\pm \gamma_eB^{i}_{NV}$, where $B_{NV}$ is the projection of the magnetic field along the NV axis.
Since the crystalline orientation of our sample is (100), each axis $x$, $y$, and $z$ coincide with the edges of the plate.

Defining $\hat{u}_1$, $\hat{u}_2$, $\hat{u}_3$ and $\hat{u}_4$ as the unit vectors of the possibles orientations of the NV axes, the symmetry implies $\hat{u}_1+\hat{u}_2+\hat{u}_3+\hat{u}_4=0$, and if the external magnetic field is $\vec{B}=B_x\hat{i}+B_y\hat{j}+B_z\hat{k}$ in the lab frame, where $z$ is perpendicular to the diamond surface, we obtain $\vec{B}.\hat{u}_1+\vec{B}.\hat{u}_2+\vec{B}.\hat{u}_3+\vec{B}.\hat{u}_4=0$. 
This defines a linear system of equations relating the vector components of the field. We solve the overdetermined linear system using a least-square minimization algorithm to obtain the best estimate for the vector components of $\vec{B}$, at each pixel.
Fig. \ref{Bwire} displays the reconstruction of a magnetic field produced by a current of 15 mA in the test DC wire. 

\begin{figure}
\centerline{\includegraphics[scale=0.58]{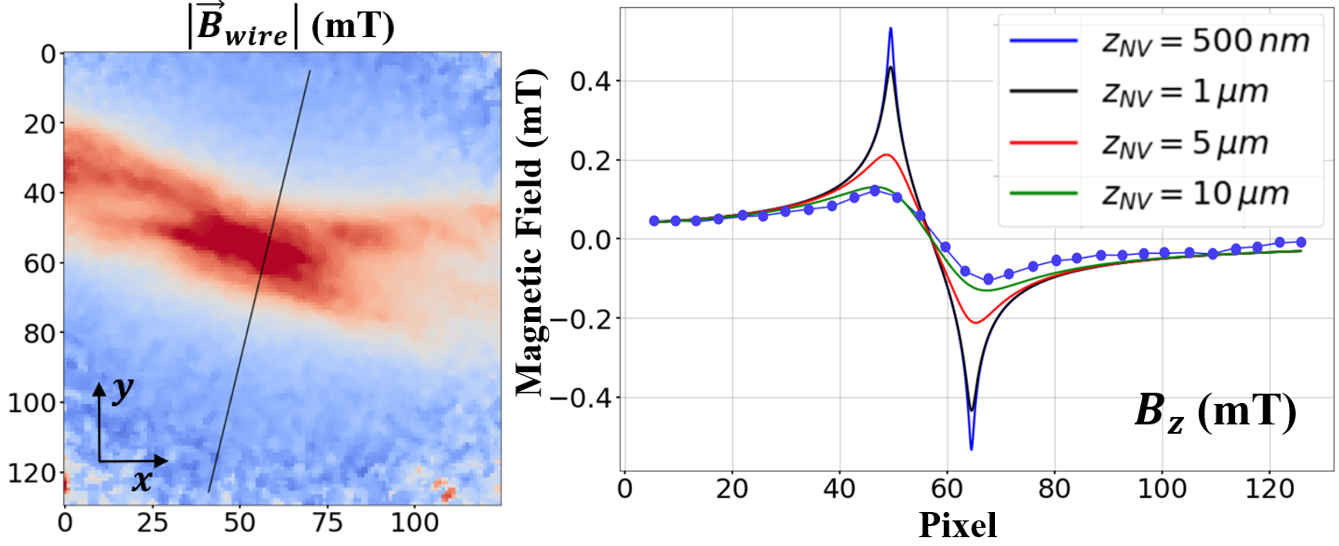}}
\caption{Points indicate de measured components of the magnetic field produced by a current $I=15\,\operatorname{mA}$. The solid color-coded lines are Biot-Savart simulations of the field produced at the NV layer for a wire of $100\,\operatorname{nm}$ thickness and width $24\,\operatorname{\mu m}$ at various probing distances $z_{NV}$. The points (blue dots) were sampled along the indicated black dashed-line, perpendicular to the wire. In the image, 1 pixel = 1.4 $\operatorname{\mu m}$. \cite{Charlie_PhD-Thesis}} 
\vspace{-1mm}
\label{Tmodel}
\end{figure}

Considering an idealized homogeneous ultra-thin (100 nm) wire of width $W$ carrying a current $I$, we can use Biot-Savart law to calculate the magnetic field produced by the wire at a given probing distance $z_{NV}$. Fig. \ref{Tmodel} shows a comparison of the profiles of the calculated (simulated) $B_z$ component with experimental points measured along a line perpendicular to the wire (black dashed-line), for that component at for several distances $z_{NV} = \{0.5,\, 1,\, 5,\, 10\} \mu\mbox{m}$. This procedure was done to all the components, resulting in the value of $z_{NV} = 10\pm 2\,\mu\mbox{m}$ as the distance that best adjusts to experimental data.

\begin{figure}[ht]
\centerline{\includegraphics[scale=0.8]{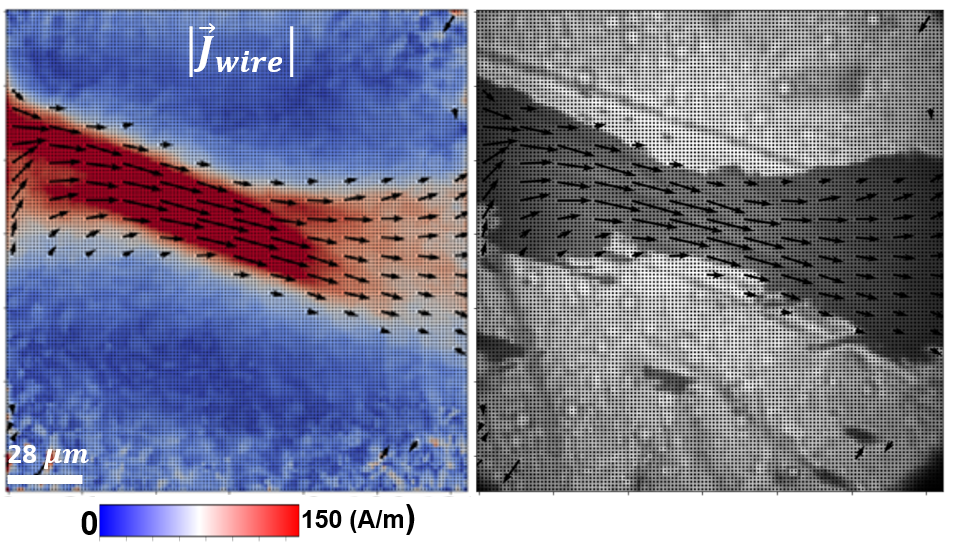}}
\caption{Calculated current density for $I=15\,\operatorname{mA}$ \cite{Charlie_PhD-Thesis}. The probe distance is assumed to be $z_{NV} = 10\,\operatorname{\mu m}$. On the right, the current density is overlaid to the microscopy image of the wire.}
\label{Jcam}
\vspace{-1mm}
\end{figure}

To calculate the density current from the measured magnetic field one assumes that the current is confined to a 2D plane \cite{Tetienne_graphene,Roth2}. In our case, the copper wire has a thickness of 100 nm and a width of around 10 $\mu$m. Under this assumption and using our measured B-field, we calculate the vector current density shown in Fig. \ref{Jcam}, for a section of the copper (DC circuit) wire. The figure also shows the vector lines of $\vec{J}$, superposed to its magnitude map, as well as a direct (bright field) microscopy image of the same portion of the copper wire.

\bigskip\bigskip
\section{Conclusion}
In our system, the sensitivity associated to a $1\,\operatorname{\mu m}^2$ area of the NV layer is $\eta = 6\,\operatorname{\mu m}/\sqrt{\text{Hz}}$. This value is comparable with the literature \cite{Tetienne_graphene,Dreau,Chipaux2015,Taylor} and is mainly limited by the contrast and linewidth of the ODMR resonances. Measured contrasts are in the range from $0.3\%$ to $0.6\%$ per pixel. Besides, in our case the laser power was limited to 20 mW on the sample. Higher powers are needed to excite larger areas, while reducing shot-noise. These are ways to improve all these limitations, as discussed in ref. \cite{Chipaux2015}. In addition, using a fully quantum protocol, based on the spin coherence and Ramsey MW pulses \cite{Taylor}, the sensitivity can be further improved, increasing the resolution to the $\operatorname{nT}/\sqrt{\text{Hz}}$ range. We are currently pursuing quantum protocols in our laboratory \cite{Lucas2021}.

An important detail in our setup is that the DC wire was on the surface of a glass cover slide and not directly on the diamond surface. Our analysis in Fig. \ref{Tmodel} shows that the distance between the NV layer and the wire was $\sim 10\,\operatorname{\mu m}$, explaining the need for larger currents. For 2D systems deposited on the top surface, near the NV layer, one can detect much lower density currents.

\section{Acknowledgment}
We thank Victor Acosta for providing the diamond sample used in this study and Eduardo de Azevêdo for helpful discussions. The authors acknowledge financial support from CAPES, CNPq, and FAPESP (grants 2019/27471-0 and 2013/07276-1).

\end{document}